\documentclass[]{ceurart}

\usepackage{listings}
\usepackage{url}
\hypersetup{
    colorlinks=true,
    linkcolor=blue,
    filecolor=magenta,      
    urlcolor=blue,
    pdfpagemode=FullScreen,
    }
\usepackage{tabularx}
\newcolumntype{t}{>{\hsize=23\hsize}X}
\newcolumntype{a}{>{\hsize=4\hsize}X}

\usepackage{subcaption}
\usepackage{todonotes}

\begin{document}

\copyrightyear{2025}
\copyrightclause{Copyright for this report is held by its authors.
  Use is permitted under Creative Commons License Attribution 4.0
  International (CC BY 4.0).}

\conference{ }

\author[1]{Luca Rossetto}[%
    orcid=0000-0002-5389-9465,
    email=luca.rossetto@dcu.ie]

\author[2]{Klaus Schoeffmann}[%
    orcid=0000-0002-9218-1704,
    email=ks@itec.aau.at]

\author[1]{Cathal Gurrin}[%
    orcid=0000-0002-9218-1704,
    email=cathal.gurrin@dcu.ie]  

\author[3]{Jakub Lokoč}[%
    orcid=0000-0002-3558-4144,
    email=jakub.lokoc@matfyz.cuni.cz] 

\author[4]{Werner Bailer}[%
    orcid=0000-0003-2442-4900,
    email=werner.bailer@joanneum.at]

\address[1]{Adapt Centre, School of Computing, Dublin City University, Dublin, Ireland}
\address[2]{Institute of Information Technology, Klagenfurt University, Austria}
\address[3]{Department of Software Engineering, Charles University, Prague, Czechia}
\address[4]{DIGITAL -- Institute for Digital Technologies, JOANNEUM RESEARCH, Graz, Austria}

\begin{abstract}
    This report presents the results of the 14\textsuperscript{th} Video Browser Showdown, held at the 2025 International Conference on Multimedia Modeling on the 8\textsuperscript{th} of January 2025 in Nara, Japan.
\end{abstract}

\title{Results of the 2025 Video Browser Showdown}

\maketitle

\section{Introduction}

The Video Browser Showdown (VBS) is an annual interactive video retrieval competition held at the International Conference on MultiMedia Modeling (MMM).
This report presents the results of the 14\textsuperscript{th} edition of this international competition (VBS 2025), which took place on the 8\textsuperscript{th} of January 2025 in Nara, Japan.
Similarly to the 2024 report~\cite{rossetto2024results2024videobrowser}, this report provides an overview of the tasks and teams and their results, without discussing them in-depth.
See the detailed analysis papers from 2023~\cite{DBLP:journals/access/VadicamoABCGHLLLMMNPRSSSTV24}, 2022~\cite{DBLP:journals/mms/LokocABDGMMNPRSSSKSVV23}, 2021~\cite{DBLP:journals/ijmir/HellerGBG0LLMPR22}, etc., for a more detailed discussion of past instances of the VBS.
This report is based on the data generated by the Distributed Retrieval Evaluation Server~\cite{DRES-TOMM}, the central VBS coordination and evaluation infrastructure.
The raw data is available from the VBS archive.\footnote{\url{https://github.com/lucaro/VBS-Archive}}

\section{Competition setup}

Building on the datasets from previous years~\cite{DBLP:journals/access/VadicamoABCGHLLLMMNPRSSSTV24}, VBS 2025 further scaled up the large and challenging datasets being used. For the first time, the complete V3C collection~\cite{RossettoSAB19} with $3\,800$ hours of heterogeneous video content from $28\,450$ video files downloaded from Vimeo, and the extended version 2 of the MVK dataset \cite{MVK} with challenging  homogeneous video content from $1\,372$ (MVK1) and $1\,820$ (MVK2) short marine videos, totaling over 28 hours of combined duration.
As already in 2024, these two datasets were complemented by another highly challenging dataset from the medical domain with about 100 hours of video content from 72 gynecological laparoscopies.

\begin{figure}[t]
    \centering
    \includegraphics[width=0.95\linewidth]{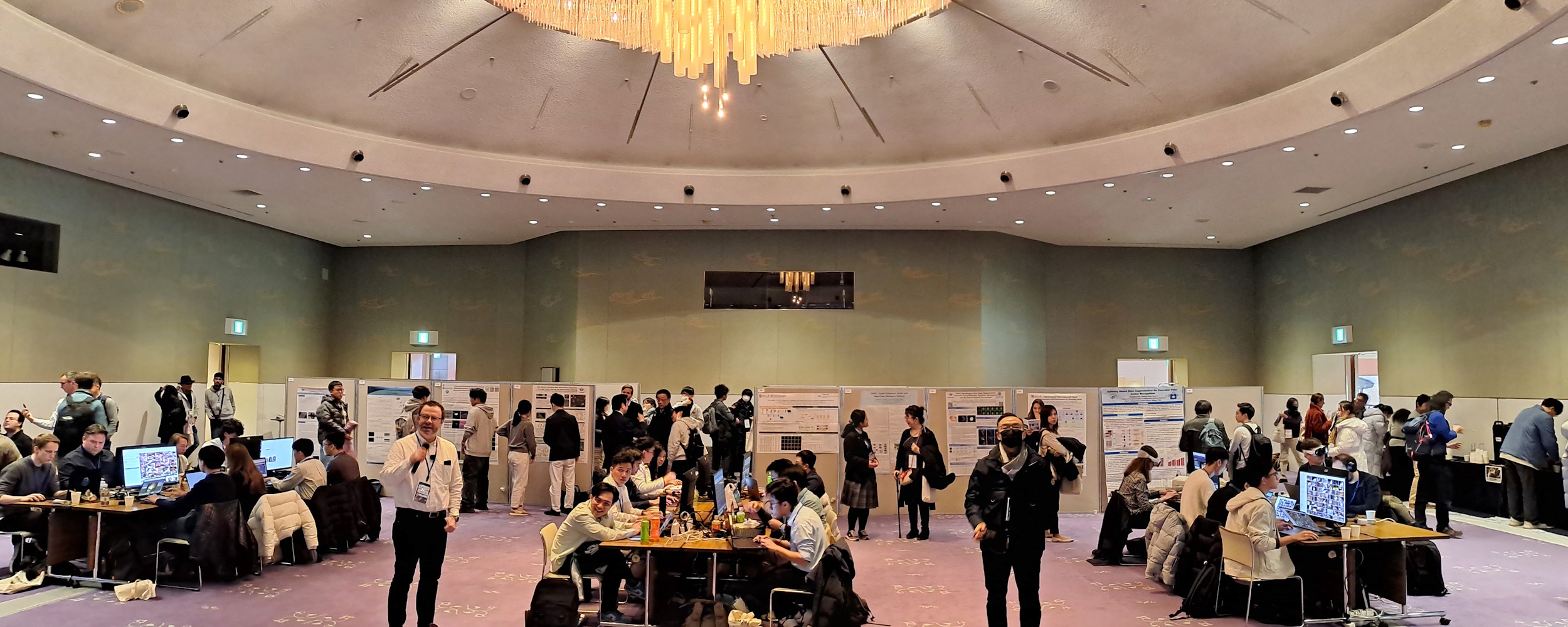}
    \caption{Setup of VBS 2025}
    \label{fig:vbs2025setup}
\end{figure}

Expert users solved four different task types: ad-hoc video search (AVS), question-answering (QAS), and known-item search (KIS) tasks presented with a visual hint (KISV) or -- as a new task type introduced in 2025 -- as incremental textual information provided in an interactive conversation (KISC).
Visual KIS tasks were presented as a scene played in a loop, while during AVS tasks, a text describing the searched concepts was presented.
QAS tasks were presented as a combination of textual description and video clips (e.g., ``\textit{How many cups of sour cream did they get?}'').

For KISC tasks, a text describing the searched scene was displayed, with additional details being added after 60 and 120 seconds into the task. The list of task types, their duration and presentation form is presented in Table~\ref{tab:taskTypes}. 

These different tasks cover different combinations of properties such as search need presentation, number of relevant requested items, and search need presentation quality (i.e., level of detail) in the space of possible task categories~\cite{DBLP:conf/mmm/LokocBBGHJPRSVV22}.

Unlike expert users, novice users worked on only three task types, including AVS and KISV as well as known-item search tasks with textual hints (KIST).

KISC has been introduced as a new task type in 2025.
It can be considered and evolution of the KIST task, which has from its beginning been designed to model the situation of someone describing the need for video content, possibly involving the inaccuracies of fading memory of content seen long time ago.
The additional hints with more details provided in KIST after 60 and 120 seconds model the aspect of recollecting more information when seeing first results, or answering (hypothetical) questions from the searcher.
An even more extreme case of this incremental information reveal as simulated recollection can be seen in the analogous task type in the Lifelog Search Challenge~\cite{DBLP:journals/access/TranNDHSLPNKDJRYAATHSG23}, where more information is made available to participants in 30-second intervals over the course of three minutes.

The increasing interest in conversational search systems 
(cf.~\cite{schneider2024engineering}), fueled by users being familiar with chat interactions with large language models, brought us to design the KISC task type.
The information is provided by an ``oracle panel'' composed of three senior judges.
Five to ten minutes before the planned start of the task, they watch a video clip that is the target of a KIST task for 100 seconds in a loop using the evaluation server (DRES)~\cite{DRES-TOMM}.
One of them prepares a short summary (the length was between 13 and 36 words at VBS 2025) and tells the summary to the participants in the room at the start of the task.
At the same time, this summary is posted in the chat of the conferencing system (accessible to remote participants and those in the room who chose to connect).
Participants may ask questions about the content orally in the room (taken by a moderator moving around in the room) or type them in the chat.
The oracle panel members answer each of the questions orally and via chat (typed by themselves or another member while or just after answering orally).
Thus all information provided as answer to a question becomes available to all the participants at the same time, refining the query with more detailed information.
In contrast to the incremental information in KIST, the information provided in KISC is guided by the questions of the participants, and thus by the content they receive in response to the query so far, using their respective systems.
While this task type serves well to model conversational search, and makes the VBS setting even more interactive, it has to be noted that it is less comparable when run in different sessions, as the questions and thus the additional hints provided may be completely different. 

For all task types, the tasks were presented through a projector and large displays directly connected to the evaluation server (DRES)~\cite{DRES-TOMM}.
Teams were grouped around multiple tables in the room with view on the projection screen or large TV monitors at the sides, so that they could see the evaluation server screen, but less so the system and solutions of other teams, see Figure~\ref{fig:vbs2025setup}.

Due to the huge dataset, the ground-truth of AVS tasks with multiple answers cannot be pre-annotated and automatically checked.
In addition, answers to QAS tasks are often in different languages or can be correctly expressed in several ways.
Therefore, VBS 2025 again used live judges who manually assessed submissions that could not be automatically checked during the competition, analogously to previous years.
Eight researchers served as live judges during at least one of the sessions.

\begin{table}[]
    \centering
    \begin{tabular}{l|l|l}
        type & duration & task presentation \\
        \hline
        visual known-item search (KISV) & 5 minutes & scene in the loop  \\
        textual known-item search (KIST) & 7 minutes & text revealed incrementally  \\
        conversational known-item search (KISC) & 7 minutes & scene described interactively \\
        ad-hoc video search (AVS) & 5 minutes & short text description \\
        questions answering (QAS) & 5 minutes & scene in the loop and text question 
    \end{tabular}
    \caption{Task types at VBS 2025}
    \label{tab:taskTypes}
\end{table}

\pagebreak
\section{Teams}

In 2025, 17 teams participated in VBS. Similar to 2024, team members were scored individually rather than in aggregate.
The 17 teams were represented by 37 individual participants.
The following lists the participating teams, ordered by their final rank in the evaluation.

\begin{enumerate}

    \item NII-UIT~\cite{DBLP:conf/mmm/GiaKTDLDMNLS25}
    \item PraK Tool V3~\cite{DBLP:conf/mmm/StrohKVVBJHL25}
    \item diveXplore~\cite{DBLP:conf/mmm/LeopoldS25}
    \item Exquisitor~\cite{DBLP:conf/mmm/SharmaKRJ25}
    \item VIREO~\cite{DBLP:conf/mmm/ChengWMHWN25}
    \item SnapSeek 2.0~\cite{DBLP:conf/mmm/HoLeHDLVNGT25}
    \item VERGE~\cite{DBLP:conf/mmm/PantelidisGPGASMGGVMK25}
    \item VEAGLE~\cite{DBLP:conf/mmm/NguyenHoHKTNHG25} 
    \item ViewsInsight 2.0~\cite{DBLP:conf/mmm/VuongHNTHLPNGT25}
    \item IMSearch 2.0~\cite{DBLP:conf/mmm/LuuQNBDLNT25}
    \item VideoEase~\cite{DBLP:conf/mmm/TranNJG25}
    \item ViFi~\cite{DBLP:conf/mmm/QuanNT25}
    \item vitrivr~\cite{DBLP:conf/mmm/RossettoG25}
    \item vitrivr-VR~\cite{DBLP:conf/mmm/SpiessRS25}
    \item MediaMix~\cite{DBLP:conf/mmm/ArnoldKWS25}
    \item PoliEste (Fusionista)~\cite{DBLP:conf/mmm/LeTDQTNN25}
    \item HORUS~\cite{DBLP:conf/mmm/NguyenAPVQTLN25}    
\end{enumerate}

\section{Results}

This section illustrates the scores, including their development over time, and some general properties of the submissions made by all participants during the evaluation.
The evaluation was again split into two parts: first, the developers of the participating systems operated their systems themselves in an \emph{expert} session. Afterwards, additional retrieval tasks were conducted with \emph{novice} participants who had not previously seen the systems they were using.
The task types between the two sessions were not identical.

\subsection{Scores}

\begin{figure}[!ht]
    \centering
    \includegraphics[width=0.925\linewidth]{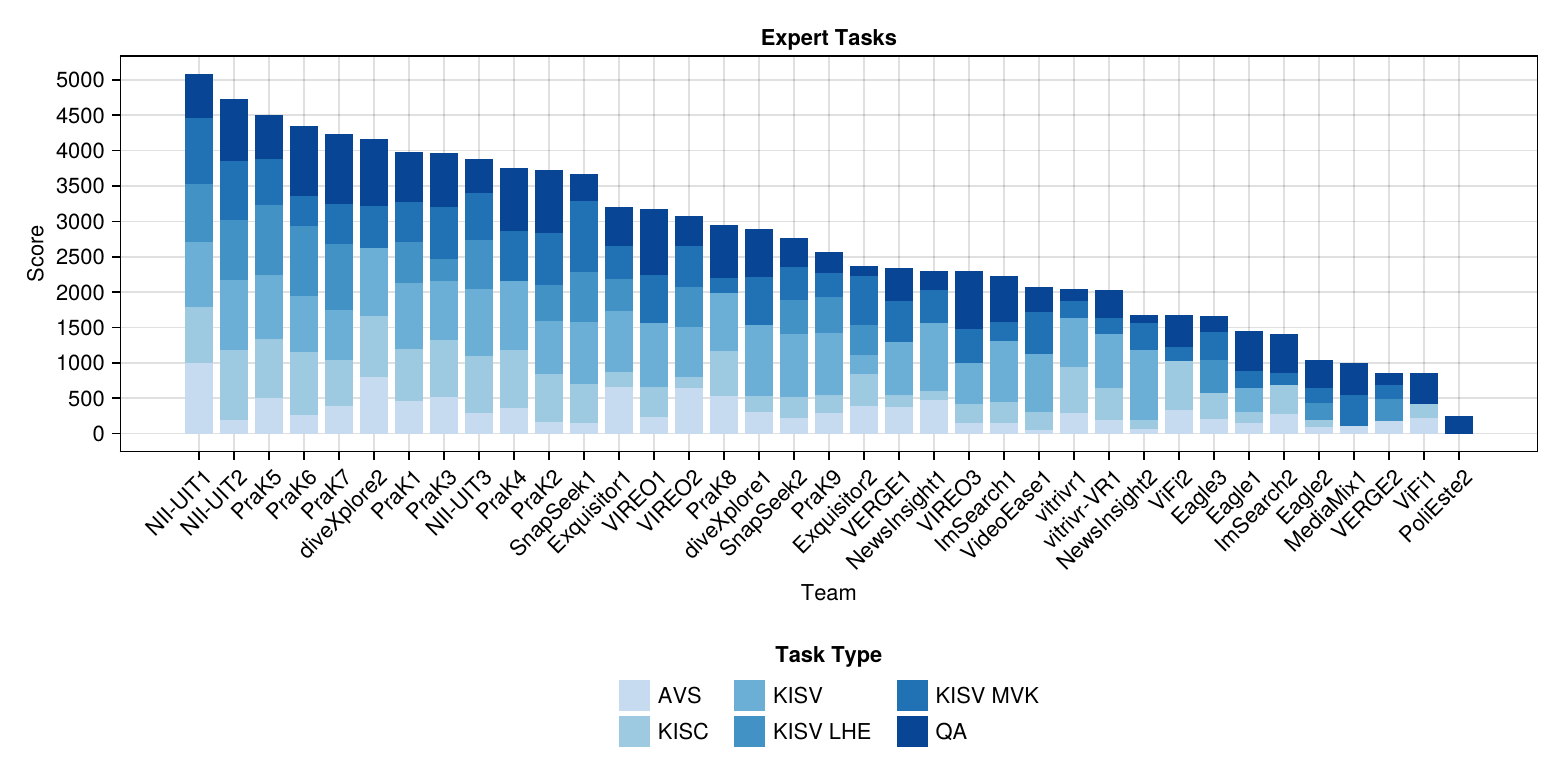}
    \caption{Scores of the expert tasks grouped by participant}
    \label{fig:scores-expert}
\end{figure}

Figure~\ref{fig:scores-expert} shows the scores of the expert tasks per participant for the four different task types: ad-hoc video search (AVS), known-item search with conversational queries (KISC), known-item search with visual queries for the three different datasets (KISV, KISV LHE, and KISV MVK), and question answering (QAS).
Only scoring teams are shown.

\begin{figure}[!ht]
    \centering
    \includegraphics[width=0.925\linewidth]{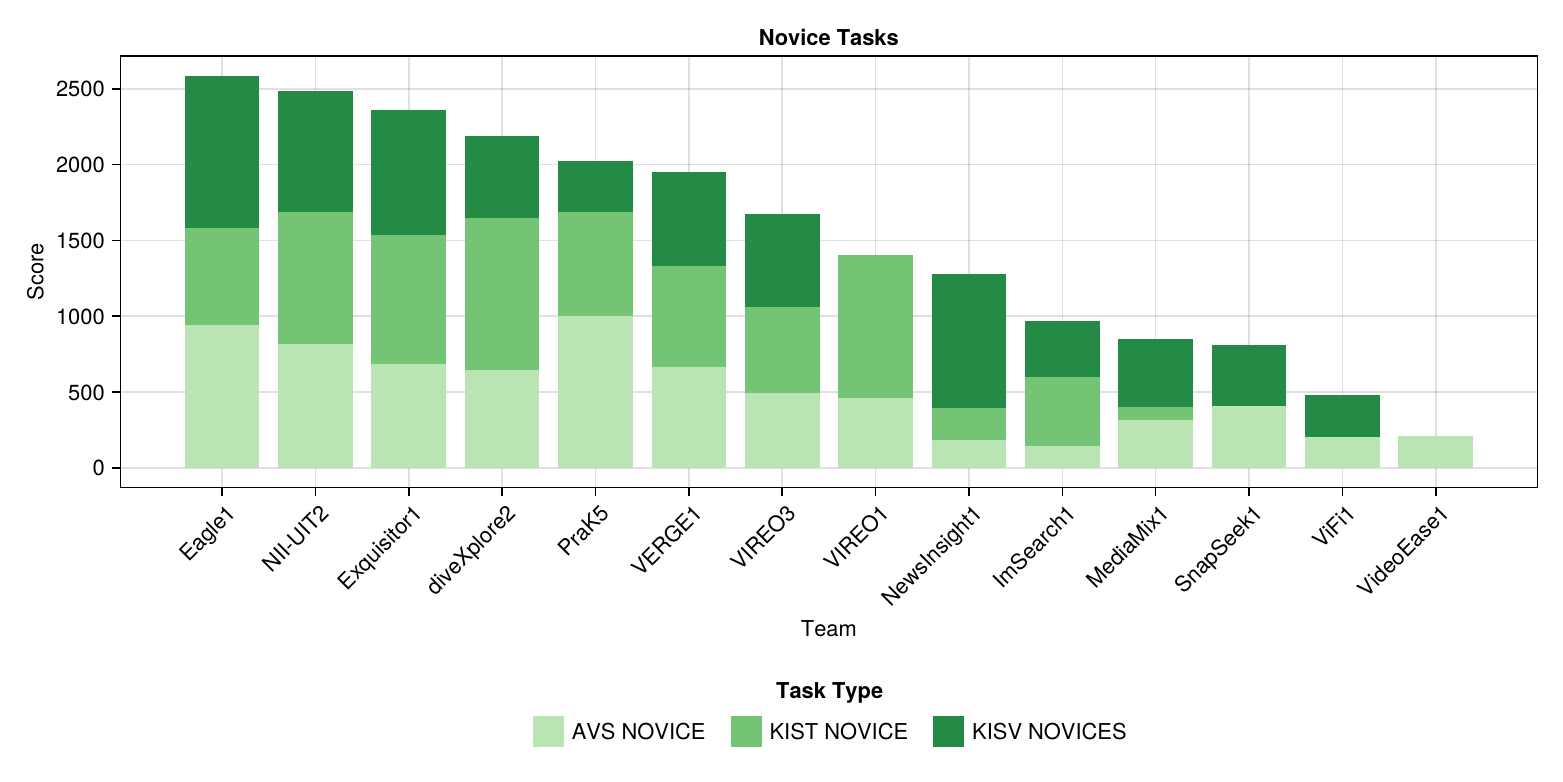}
    \caption{Scores of the novice tasks grouped by participant}
    \label{fig:scores-novice}
\end{figure}

Figure~\ref{fig:scores-novice} shows the same score distribution for the novice session.
In this session, no Known-Item Search tasks with conversational queries were performed and only the V3C dataset was used.

\begin{figure}[!ht]
    \centering
    \includegraphics[width=0.925\linewidth]{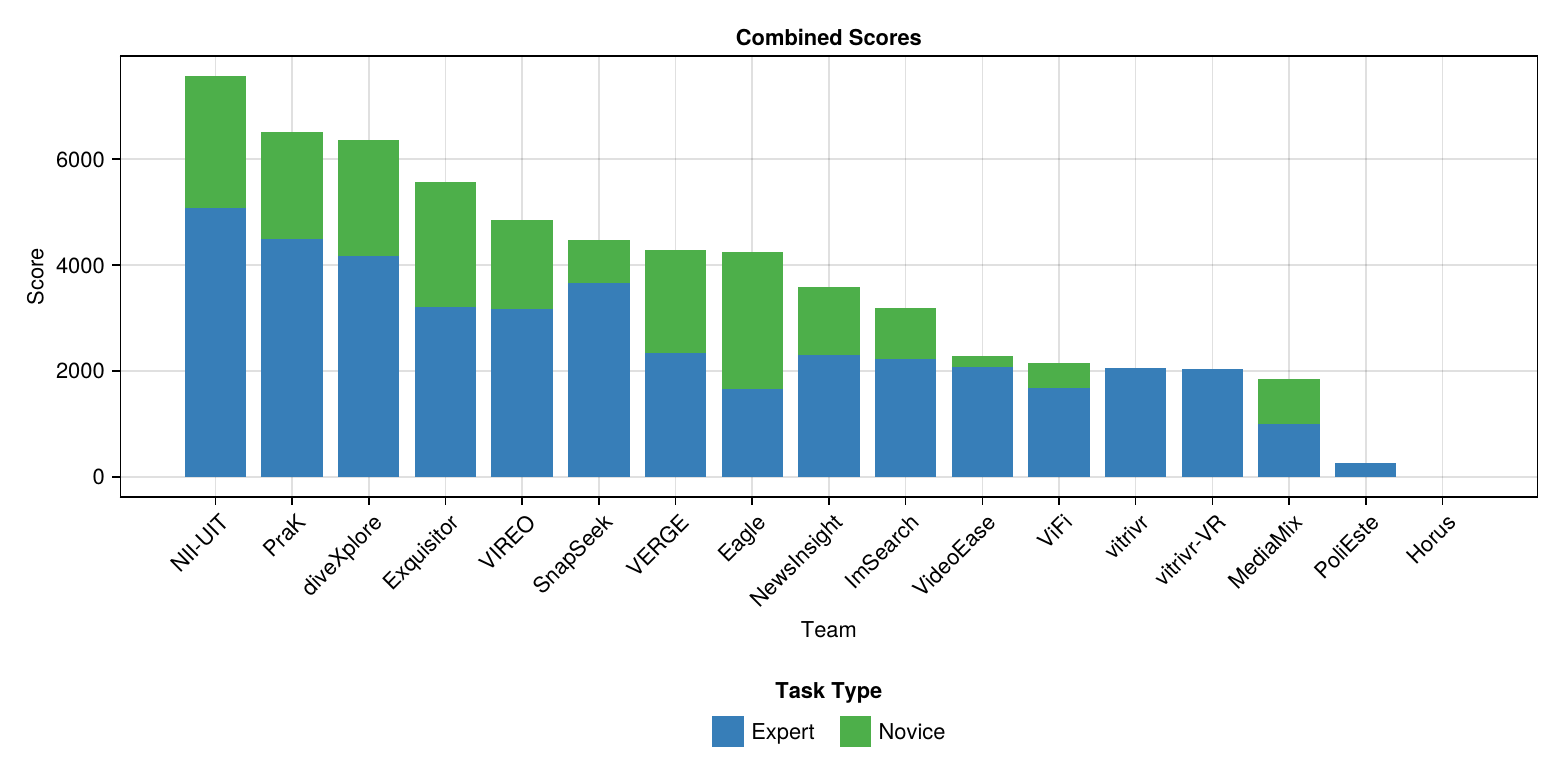}
    \caption{Combined scores, using the best-performing expert and novice per team}
    \label{fig:scores-combined}
\end{figure}

To determine the overall best-performing system/team, the highest score per system was combined from either session. The resulting final scores are shown in Figure~\ref{fig:scores-combined}.

\pagebreak

\subsection{Submissions}

\begin{figure}[!ht]
    \centering
    \includegraphics[width=\linewidth]{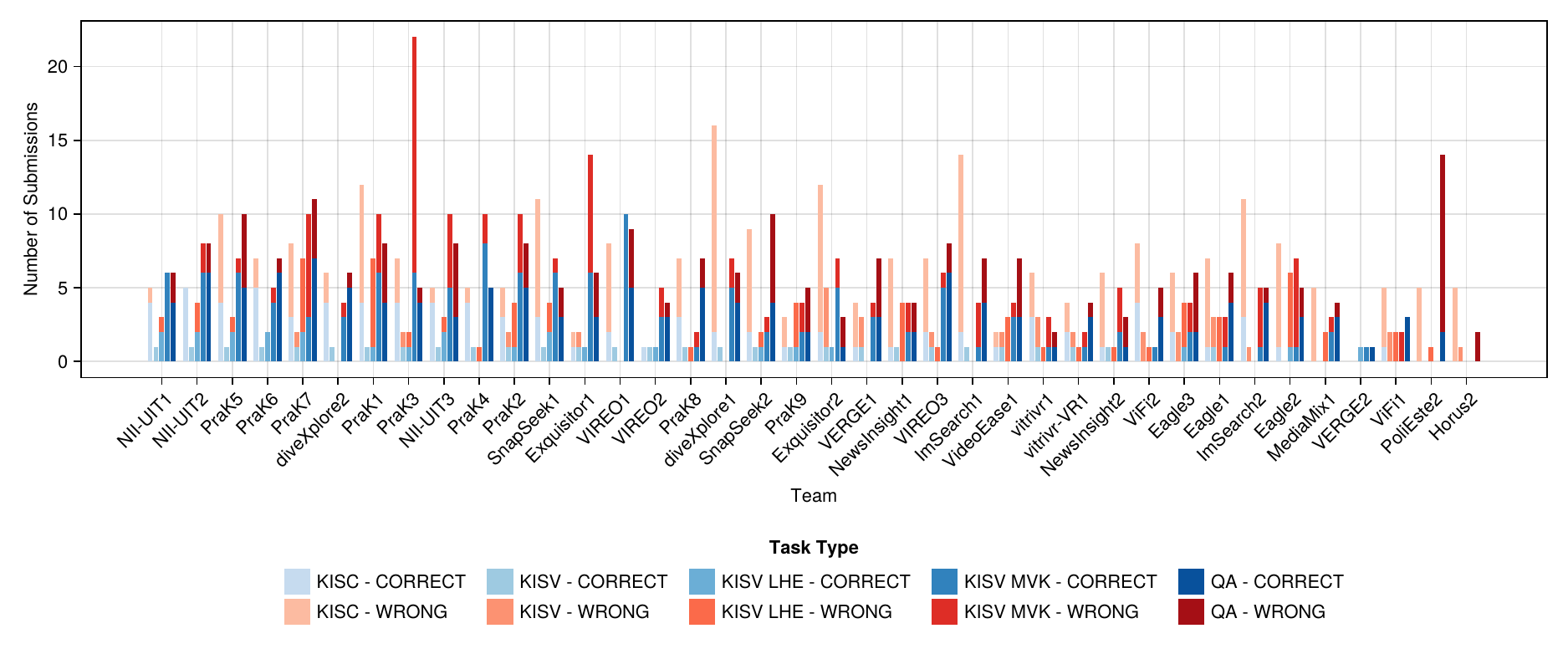}
    \caption{Number of correct and incorrect submissions per task type and participant}
    \label{fig:status}
\end{figure}

Figure~\ref{fig:status} shows the number of correct and incorrect submissions per type of task and participant. Ad-hoc search tasks are omitted due to their comparatively high number of submissions.

\begin{figure}[!ht]
    \centering
    \includegraphics[width=\linewidth]{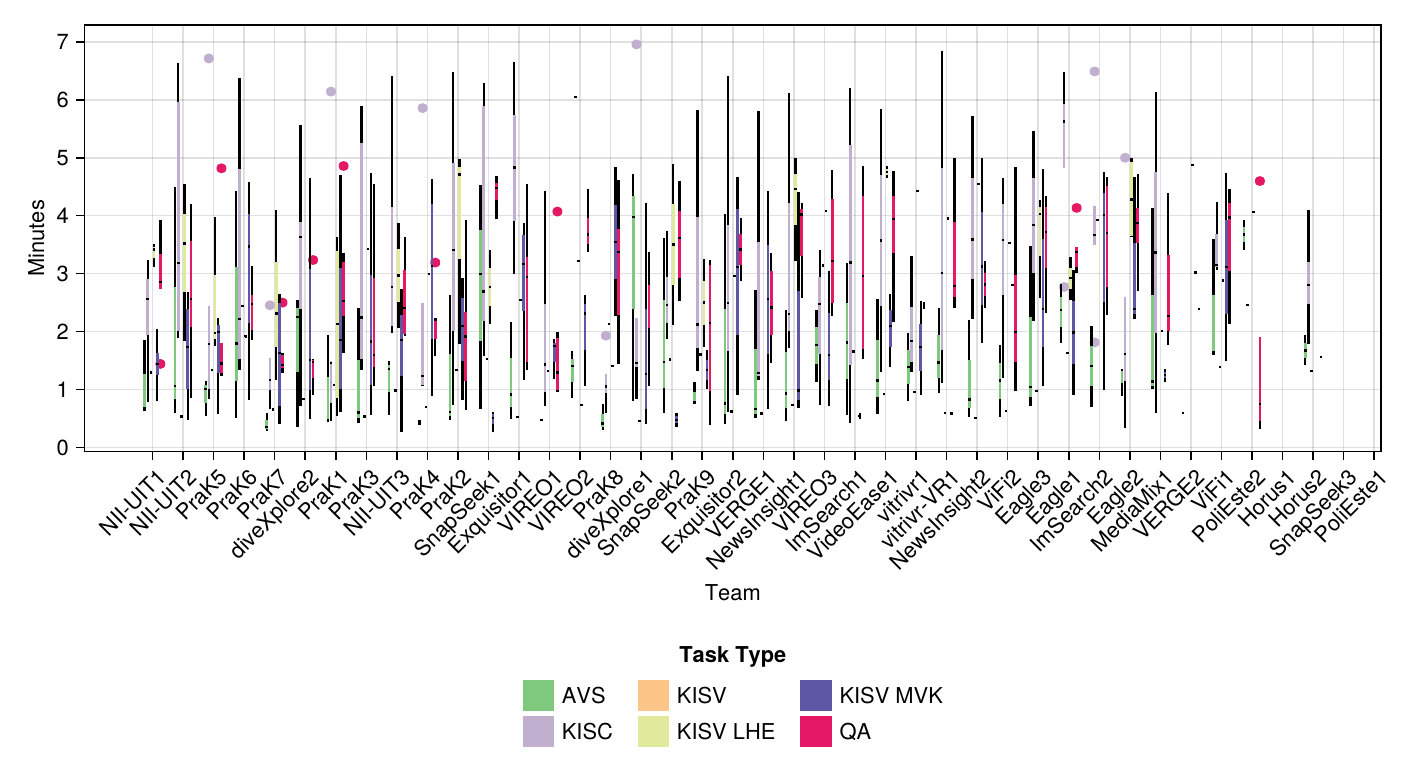}
    \caption{Distribution of the time taken (in minutes) until the first submission per task type and participant}
    \label{fig:time-to-submission}
\end{figure}

\noindent Figure~\ref{fig:time-to-submission} shows the distribution of the time taken until any participant submitted a first submission for any type of task. This serves as a rough estimate of the query processing efficiency of the different systems.
Due to the high variability in novice performance, the figure only shows times from the expert tasks.

\pagebreak
\subsection{Ranking over time}

\begin{figure}[!ht]
    \centering
    \includegraphics[width=\linewidth]{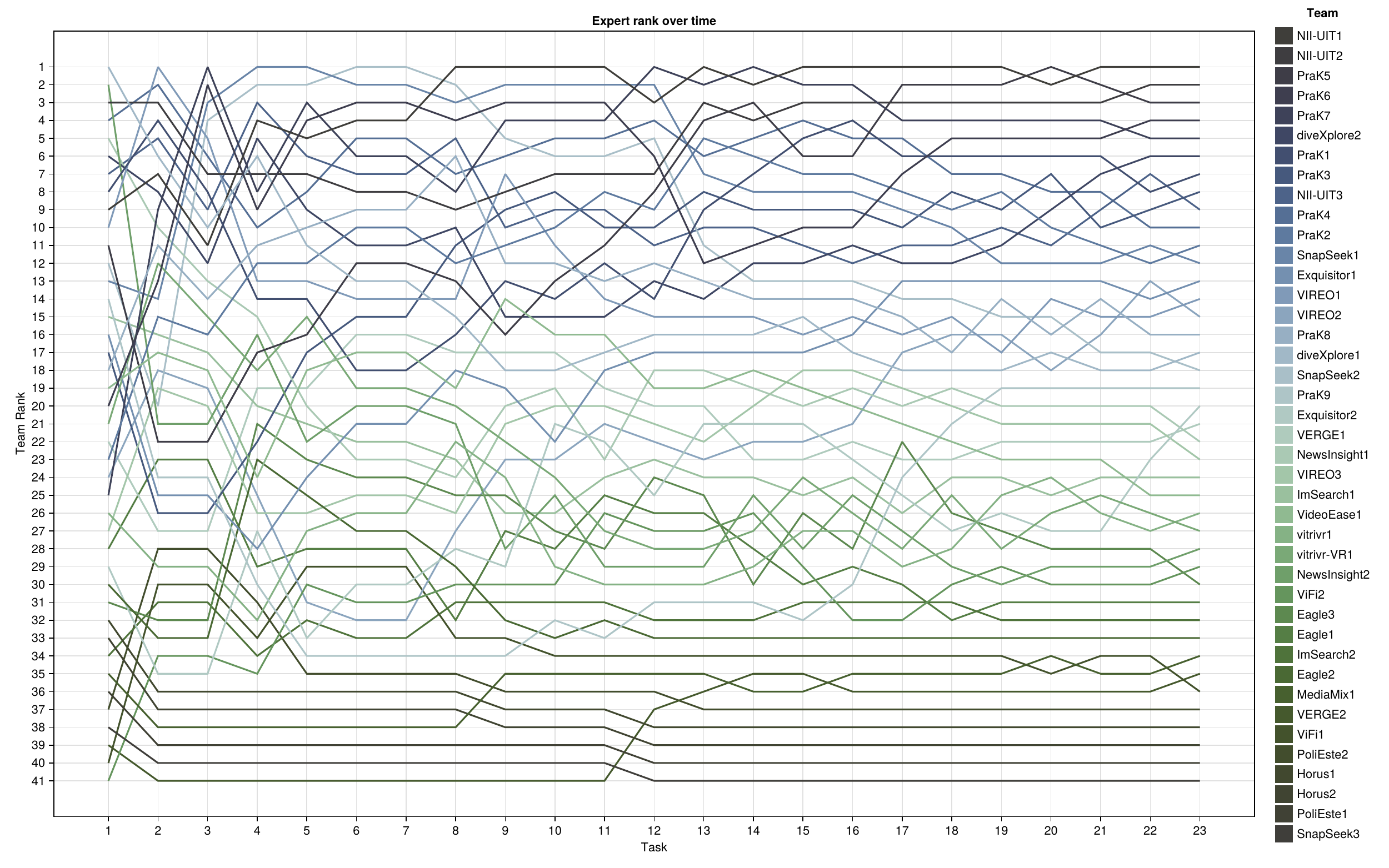}
    \caption{Raking of participants in the expert tasks over time}
    \label{fig:expert-rank-over-time}
\end{figure}

\begin{figure}[!ht]
    \centering
    \includegraphics[width=\linewidth]{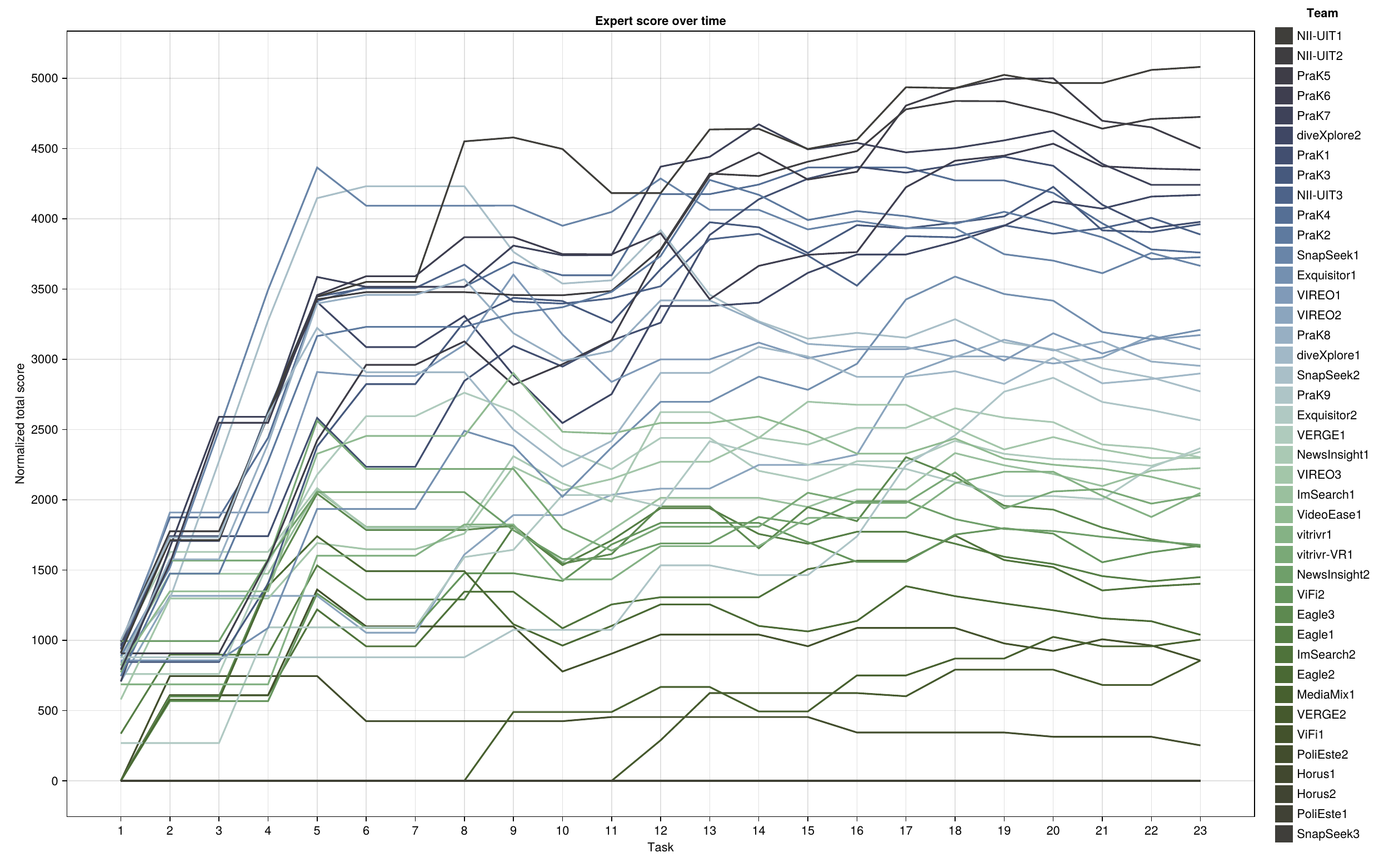}
    \caption{Score of participants in the expert tasks over time}
    \label{fig:expert-score-over-time}
\end{figure}

\noindent Figure~\ref{fig:expert-rank-over-time} shows the ranking of the expert participants after every task in the evaluation and Figure~\ref{fig:expert-score-over-time} shows the score of the expert participants after each task in the evaluation.

\pagebreak

\begin{figure}[!ht]
    \centering
    \includegraphics[width=\linewidth]{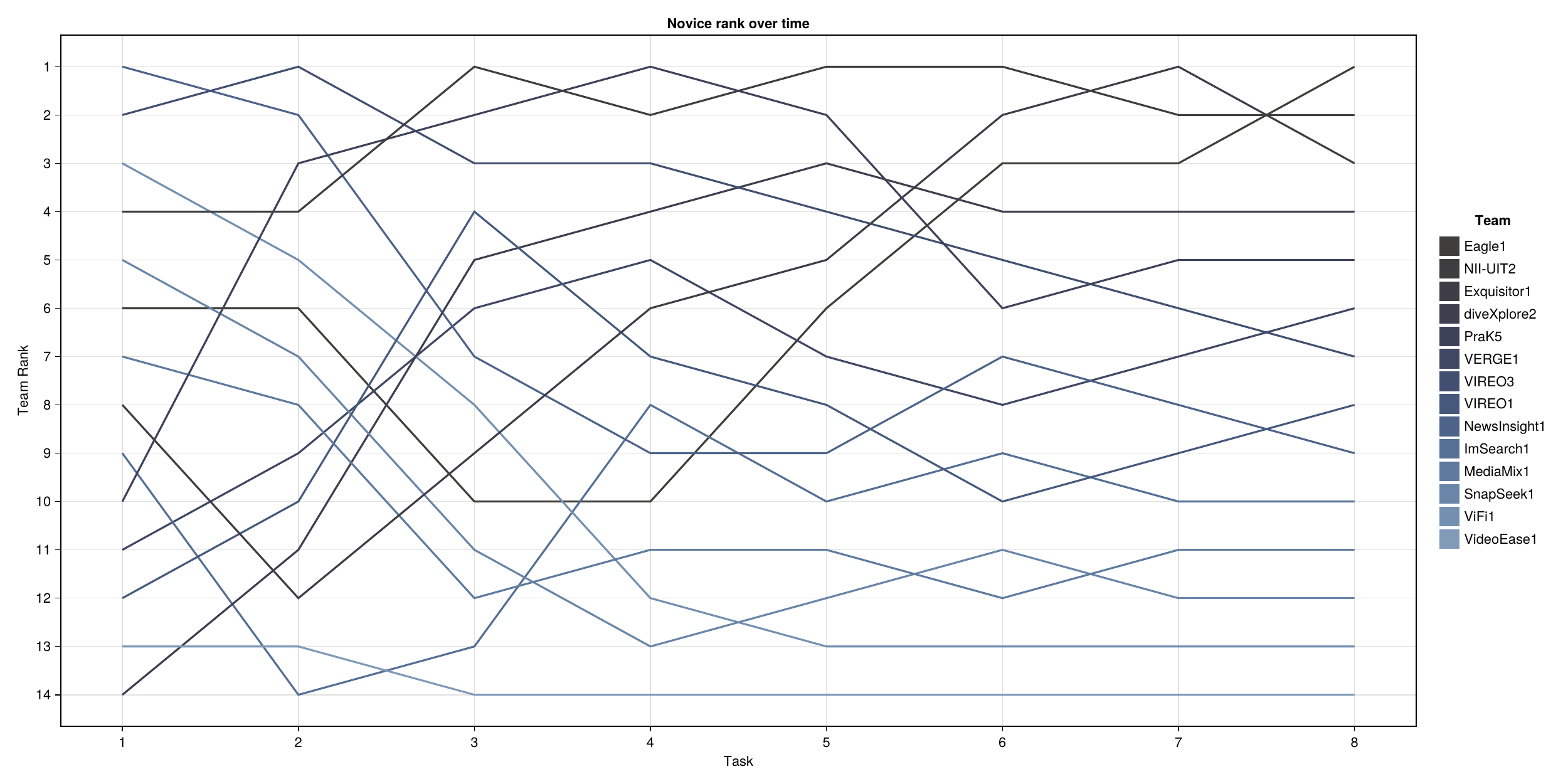}
    \caption{Raking of participants in the novice tasks over time}
    \label{fig:novice-rank-over-time}
\end{figure}

\begin{figure}[!ht]
    \centering
    \includegraphics[width=\linewidth]{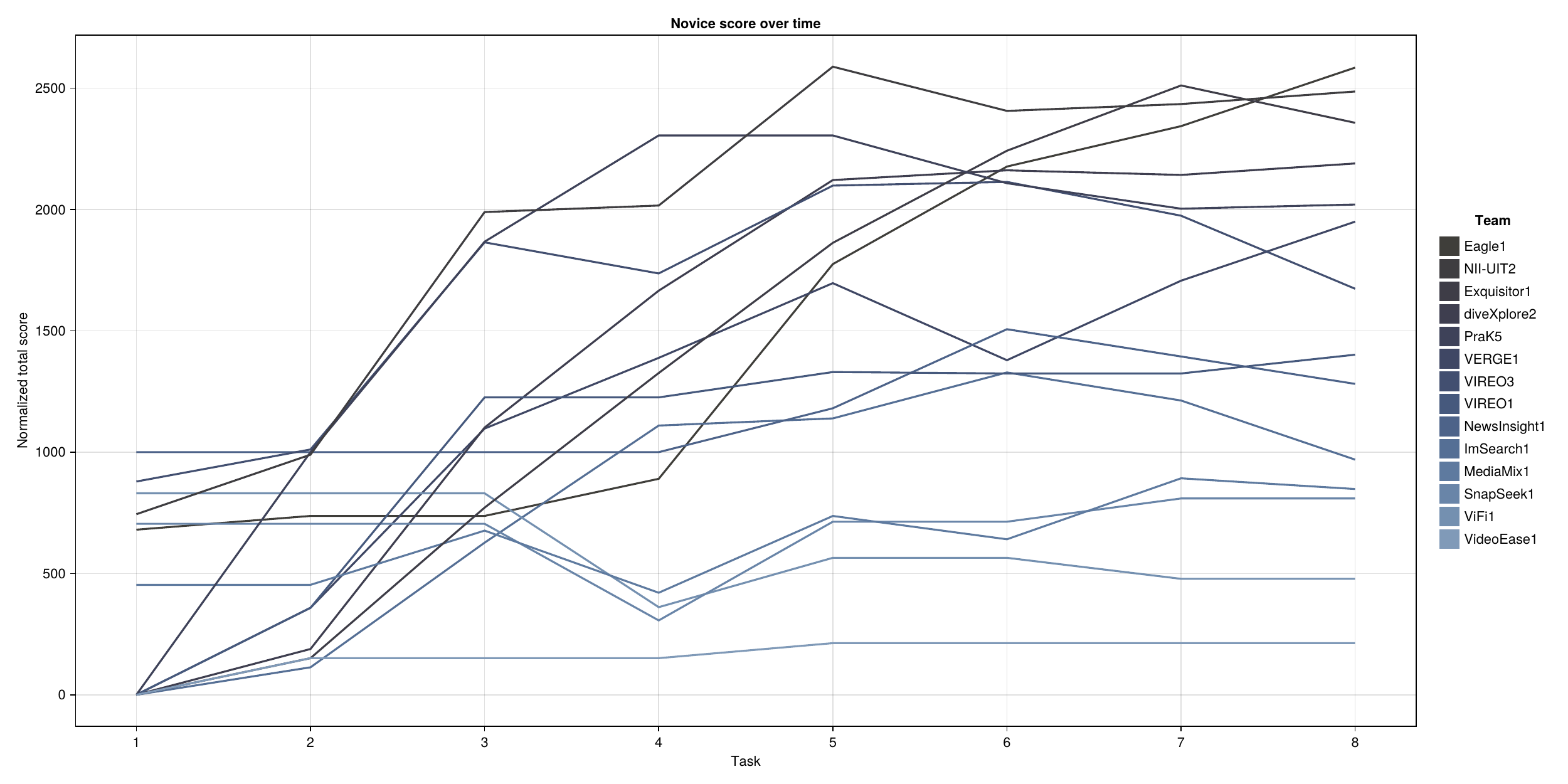}
    \caption{Score of participants in the novice tasks over time}
    \label{fig:novice-score-over-time}
\end{figure}

\noindent Figure~\ref{fig:novice-rank-over-time} shows the ranking of novice participants after each task in the evaluation and Figure~\ref{fig:novice-score-over-time} shows the score of novice participants after each task in the evaluation.

\section{Summary}
VBS 2025 showcased the latest advancements in interactive video retrieval. A total of 17 international teams participated in live content search tasks on a massive video dataset comprising the entire V3C collection, MVK 2, and LapGyn -- amounting to nearly 4,000 hours of video content. The competition not only highlighted the strong international interest in AI-based video retrieval but also demonstrated significant progress in the field. Teams successfully tackled complex search challenges with remarkable speed and reliability, including novel query types such as conversational known-item search (KISC).

\pagebreak
\bibliography{bibliography}

\end{document}